\documentclass[]{article}
\usepackage{psfig}
\usepackage{apjbib}
\usepackage{aaspp4}
\usepackage{flushrt}

\def\mathnew{\mathsurround=0pt}

\def\kms{\ {\rm km\,s^{-1}}}
\def\hmpc{\ h^{-1}{\rm Mpc}}

\def\Mvir{M_{\rm vir}}
\def\Rs{R_{\rm s}}
\def\Rvir{R_{\rm vir}}
\def\rhos{\rho_{\rm s}}

\def\simov#1#2{\lower .5pt\vbox{\baselineskip0pt \lineskip-.5pt
       \ialign{$\mathnew#1\hfil##\hfil$\crcr#2\crcr\sim\crcr}}}
\def\simgreat{\mathrel{\mathpalette\simov >}}
\def\simless{\mathrel{\mathpalette\simov <}}
\def\nbody{$N$-body}
\def\solmass{M_{\odot}}
\def\hsolmass{h^{-1}M_{\odot}}

\def\today{\ifcase\month\or
  January\or February\or March\or April\or May\or June\or
  July\or August\or September\or October\or November\or
  December\fi
  \space\number\day, \number\year}

\begin{document}
\bibliographystyle{apjbib}

\title{Young galaxies: what turns them on?}

\author{T.S. Kolatt$^1$, J.S. Bullock$^1$, R.S. Somerville$^2$,
Y. Sigad$^2$, P. Jonsson$^3$, \\
  A.V. Kravtsov$^4$, A.A. Klypin$^4$, J.R. Primack$^1$,
  S.M. Faber$^5$ \& A. Dekel$^2$ }


\vskip 2pt
{\center
 \noindent $^1$Physics Department, University of California, Santa Cruz,
     CA 95064, USA \\
\noindent $^2$Racah Institute for Physics, The Hebrew University,
     Jerusalem 91904, Israel \\
\noindent $^3$Astronomy Department, University of California,
Santa Cruz, CA 95064 USA \\
\noindent $^4$Astronomy Department, New Mexico State University,
Box 90001, Dept. 4500, Las Cruces, NM 88003, USA \\
\noindent $^5$UCO/Lick Observatory, University of California, Santa Cruz,
CA 95064 USA \\}
\vskip 2pt

\begin{abstract}
Lyman break galaxies (LBGs)
at $z \sim 3$ exhibit number densities and
clustering similar to local $L_*$ galaxies; however, their star
formation rates (SFRs) are much higher.
We explore the scenario in which LBGs are
starburst galaxies triggered by collisions, and thus provide an
explanation for these key properties.
The relative paucity of starburst galaxies at low redshift can
be attributed to a much higher collision rate in the past.
We use high-resolution cosmological $N$-body simulations and a
hierarchical halo finder to estimate the galaxy collision rate
as a function of time in the popular $\boldmath{\Lambda}$CDM
cosmological model.
We find that bright collisional starbursts are
frequent enough to account for most of the high-$z$ ($2.5-4.5$)
LBGs.
Moreover, many of the objects are of relatively small mass,
but they cluster about large-mass halos.
They therefore exhibit strong clustering, similar to that
observed and stronger than that of the relevant massive halos.
\end{abstract}

\section{Introduction}
Data from the Hubble Deep Field (reviewed in \cite{dickinson:98})
and ground-based telescopes
(\cite{steidel:96a,steidel:96b,lowenthal:97,steidel:98})
have revealed a population of galaxies at $z\simeq2.5-4.5$.  They were
found by multicolor photometry
exploiting the characteristic spectral attenuation
shortward of the Lyman limit, so
these galaxies are referred to as ``Lyman break galaxies''.
The LBGs are observed to be abundant and highly clustered, with
comoving number densities and clustering properties at $z\simeq3$
comparable to those of present-day bright ($\geq L_*$) field
galaxies (\cite{giavalisco:98,adelberger:98}).
They are forming stars at a high rate, comparable to local
``starburst'' galaxies, and are much smaller
than similarly bright galaxies nearby.

There are competing views regarding the nature of LBGs.
In one view
(\cite{steidel:96b,adelberger:98,baugh:98,governato:98} ,Mo, Mao, \& White
1999),
most LBGs are large galaxies quiescently forming stars
at the bottom of the potential wells of massive dark matter halos.
We refer to this idea as the ``central quiescent'' scenario.
An alternate view (\cite{lowenthal:97}, Somerville, Primack, \& Faber 1998)
maintains that LBGs are mainly galactic starbursts
triggered by collisions between small, gas-rich galaxies.  We refer
to this idea as the ``collisional starburst'' scenario.

Within the central quiescent scenario, there is
roughly a one-to-one relationship between the LBGs and massive
halos (\cite{steidel:96b,adelberger:98}).
This helps explain the strong clustering of observed LBGs, since
in CDM (\cite{bfpr}) theories of hierarchical structure formation,
massive objects are more clustered than low-mass objects.
It has been shown that the clustering properties of LBGs
at $z\sim3$
can be reproduced within various CDM models if the LBGs are associated
with the most massive collapsed  dark-matter (DM) halos at that
epoch (\cite{jing:98,wechsler:98,adelberger:98}).
More detailed modeling appears generally consistent with the
central quiescent framework (\cite{baugh:98,governato:98}).

Until the simulations reported here, there have been no
predictions for clustering properties and only rough estimates of number
densities of LBGs within the collisional starburst picture.
Somerville, Primack, \& Faber (1999) used
a semi-analytic treatment in order to compare the
properties of individual galaxies in the two scenarios,
and argued that the high star formation rates,
small emission line-widths (\cite{pettini:98}) ($\sim70$ km/s),
young ages (\cite{sawicki:98})
and high star formation surface densities (\cite{lowenthal:97,heckman:98})
of  LBGs are more easily explained within the collisional
starburst model.
However, there were many unanswered questions since only
simple approximations (\cite{makino-hut})
were used to estimate the merger rate of sub-halos at high redshift
$z \simgreat 3$
and there was no way to calculate the spatial distribution
of collisional starbursts.

In order to establish whether collisional starbursts are a plausible
origin for LBGs, we now
ask whether the collisions in hierarchical scenarios
can match the observed number density and clustering properties.
These questions are addressed
in this letter using high resolution $N$-body simulations.
We first present the simulations and accompanying halo finder,
and briefly explain our prescription for identifying collisions
and estimating the resulting luminosity.
The number density of the observed LBGs is then compared to the
simulated number density as a function of redshift, and the
correlation function for the simulated LBGs is presented.

\begin{figure}[t!]
{\bf Figure 1 can also be retrieved from
 http://physics.ucsc.edu/users/tsafrir/fig1.html
 (USA) or http://www.fiz.huji.ac.il/\~\,tsafrir/fig1.html (Europe)}
\caption{Part of the simulation volume at $z=2.9$.
Spheres are dark-matter halos, color-coded by mass
[the color bar is labeled with $\log(M_{\rm halo}/\hsolmass)$]
and sized in proportion to their virial radii.
Stars indicate the locations of all halo collisions
occurring at $3.9>z>2.9$, an interval of $\sim 600$ Myr.
Since each burst lasts a relatively short
time $\tau_{\rm vis} \sim 100$ Myr, only $\sim 1/6$  of
the bursts would be seen at any one time.
Most collisions occur in or near relatively massive halos,
resulting in strong large-scale clustering.
}
\label{fig:map}
\end{figure}

\section{\nbody\ simulations and halo/collision finders}
Only recently have cosmological $N$-body simulations reached the
stage where
halo sub-structure can be resolved (e.g., \cite{klypin:overcoming}).
Our simulations make use of the ART code (Kravtsov, Klypin, \& Khokhlov 1997)
which utilizes
an adaptive grid to obtain the unprecedented resolution necessary for
identifying collisions between well-resolved galactic halos (or sub-halos)
in a cosmological volume.
The simulations followed the evolution of the DM
in the popular $\Lambda$CDM
model in which the present mass density of the Universe is $\Omega_{m}=0.3$,
and a cosmological constant $\Omega_{\Lambda}=1-\Omega_m=0.7$ makes the
Universe flat. The Hubble constant is $h=0.7$
(i.e., H$_0=70 \kms$ Mpc$^{-1}$),
the primordial fluctuations are scale invariant, and the mass fluctuation
amplitude is $\sigma_8=1.0$.  Our main results are based on a simulation
run down to $z=1.7$
with particle mass $m_p = 1.3\times 10^{8} \hsolmass$ in a $30\hmpc$ box,
but we have also used another ART simulation with the
same number of particles ($256^3$) in a $60\hmpc$ box run to $z=0$.
We locate DM halos using a bound-density-maximum
halo finder (\cite{klypin:overcoming}), now extended to cope with halo
interactions and sub-structure (\cite{Bullock_thesis}).
Halos are modeled by the density profile (Navarro, Frenk, \& White 1996)
$\rho_{\tiny\tiny\rm NFW}(r)=\rhos/[(r/\Rs)(1+r/\Rs)^2]$,
which provides a characteristic radius ($\Rs$), virial radius
($\Rvir$) and mass ($\Mvir$), and the associated fitting errors.
The profile of a sub-halo may be truncated short of its $\Rvir$.
We treat only halos with $\Mvir > 7 \times10^{9}\hsolmass$
($>50$ particles).
This is sufficient to resolve collisional starburst
LBGs according to the luminosities assigned by our prescription
(see below).
Halo collisions are identified
using pairs of stored simulation
outputs at redshifts $z_2 > z_1$. For each halo at $z_1$, we search for
sets of particles that originated in different halos at $z_2$.
If the centers of two such sets overlap within their $\Rs$
radii at $z_1$, a collision is declared (\cite{collision_rate}).
Figure~\ref{fig:map} shows the locations of such collisions.

\begin{figure}[t!]
\centerline{\psfig{file=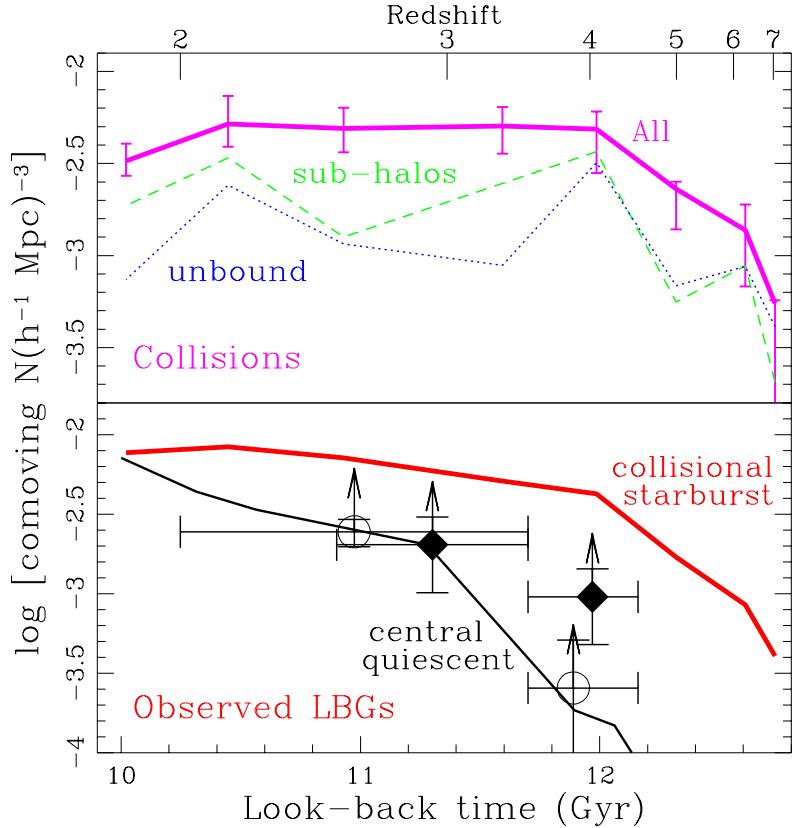,height=11truecm,width=11truecm}}
\caption{
Comoving number density of
collisions and observed LBGs as a function of look-back time.
The solid line in the top panel refers to
halo collisions in an $N$-body simulation of $\Lambda$CDM.
Here collisions between halo with masses $> 7\times10^{9} \hsolmass$
are assumed to be visible for a constant duration
$\tau_{\rm vis} = 100$ Myr.
Errors are estimated from the collision-identification procedure.
The dashed curve is the partial contribution ($\sim 50\%$) from
collisions involving at least one sub-halo
and the dotted line is the contribution of unbound collisions.
The bottom panel compares the number of LBGs actually
observed with predictions.  The thick solid curve represents
collisional starbursts brighter than
apparent magnitude ${\cal M}_{\rm AB}=25.5$,
corresponding to rest-frame $1600$~\AA.
Observational estimates for LBGs (${\cal M}_{\rm AB} < 25.5$)
from the Hubble Deep Field  (Pozzetti et al. 1998) 
and newer ground-based observations
with a much larger survey volume (Adelberger et al. 1998, Steidel
et al. 1998) 
are depicted by the open circles and solid diamonds respectively.
The upward arrows on the data are an approximate correction for
inferred dust absorption in the LBGs (see text).
The thin solid line is estimated for the central
quiescent scenario (see text).
}
\label{fig:n_lbg}
\end{figure}

\section{From collisions to bursts}

We assume that each small DM halo
at $z\sim 3$ contains a gas-rich galaxy at its center and that
each collision results in a starburst.
Mergers are included not only between isolated halos, but also
between ``sub-halos'' ---
halos that reside within the virial radius of larger halos.
Note that the analytic predictions for merger rates obtained by
Press-Schechter
approximations (\cite{lc:93}) are not sufficient here because they
miss sub-halos of isolated systems, they
are limited in predicting the progenitor
mass spectrum (\cite{sk:99}), and they ignore spatial correlations.

The top panel in Figure~\ref{fig:n_lbg} shows the time-evolution
of the number density of identified collisions as a function of
look-back time,
assuming a duration of visibility $\tau_{\rm vis}=100$ Myr.
About one half of the collisions at $z\simless 4$ involve sub-halos;
such collisions would have been missed without these very high-resolution
N-body simulations, which significantly supersede the cruder treatment
of collisions in the semi-analytic investigation~ (\cite{spf:98}).
Unbound collisions
(those in which the two halos are not subsequently bound)
are not accounted for at all in the semi-analytic
models.  Also, the dynamical friction time scale used in SAMs tends to
over-estimate the host -- subhalo collision time scale
(\cite{collision_rate}), the subhalo -- subhalo collision
approximation (\cite{makino-hut}) is somewhat simplistic, and the progenitor
mass spectrum is uncertain (\cite{sk:99,slkd:99}).
The collision rate per \textit{physical volume} (not shown) declines
$\propto (1+z)^3$ for $z\simless2.5$, in general agreement
with theory (\cite{lc:93}) and observations.
This was determined for halos $> 10^{11} \hsolmass$
using our simulation in a larger volume with lower mass resolution.

To assess how many collisions
should actually be
observable, luminosities are assigned as follows.
We assume that before a collision each galaxy has a cold-gas
reservoir $m_{\rm g} = f_{\rm g} f_{\rm b} m_{\rm halo}$,
where $f_{\rm b}$ is the fraction of mass in baryons
($f_{\rm b} \equiv \Omega_{\rm b} \Omega^{-1}_{\rm m}$)
and $f_{\rm g}$ is the fraction of baryons in cold gas.
We assume $\Omega_{\rm b} = 0.018 h^{-2}$ and $f_{\rm g}=0.3$.
Based on simulations including gas dynamics and star
formation (Mihos \& Hernquist 1994a, 1994b, and our new simulations at
$z\sim 3$ with $f_{\rm g}\sim 0.3$ using an updated version of the same code),
we divide the collisions into major ($m_2/m_1 > 0.25$) and
minor collisions, and assume that during a burst of duration
$\tau_{\rm burst}= 50$ Myr, 75\%
and $50\%$ of the gas is converted into stars respectively.
Gas depletion due to multiple collisions
is ignored because we find that at most $4\%$ of the matter in
colliding halos at $z \geq 2$ has participated in a previous encounter.
We estimate the apparent magnitude of collisional starbursts
${\cal M}_{\rm AB}$ in the band equivalent to $1600$~\AA\ rest-frame,
and $\tau_{\rm vis}$,
using Bruzual-Charlot (GISSEL98) stellar-population synthesis
models (assuming solar metallicity and a Salpeter initial
mass function).
The upper heavy line in the
bottom panel of Figure~\ref{fig:n_lbg} shows the time-evolution
of the number density of observable
LBGs with ${\cal M}_{\rm AB} < 25.5$,
for the collisional starburst model.
At $z\sim 3$, only burst events involving halos $>8 \times
10^9 \solmass$ contribute to the population of ${\cal M}_{\rm AB} < 25.5$
galaxies
($>4 \times 10^9 \solmass$ at $z\sim2$).
Because of compensating effects, varying
both $\tau_{\rm burst}$ and star formation efficiency
by factors of 2 either way results in changes $< 30\%$ in
observable
number density.
The {\it observed\,} number densities
of LBGs brighter than ${\cal M}_{\rm AB}=25.5$ are shown for comparison,
calculated from the latest data
(\cite{pozzetti:98,adelberger:98,steidel:98})
assuming the simulated cosmology.
The predicted number densities are somewhat larger than those observed,
and thus allow for dust extinction, which we have not included.
The arrows on the data points result from assuming a (conservative)
factor of three in dust extinction, coupled with
the $z\sim3$ and $z\sim4$ luminosity functions estimated
by recent ground observations~ (\cite{steidel:98}).

For comparison, we show the predictions of a central quiescent model.
We assume that every sufficiently massive halo hosts one LBG whose
luminosity is tightly correlated with the halo mass,
and we obtain an effective (constant) mass-to-light ratio ($M/L$) by
adjusting
the halo mass threshold at $z=3$ ($M\simeq8\times10^{11}\hsolmass$)
to reproduce the observed abundance of LBGs
(\cite{steidel:98}). We then predict the density evolution
assuming that $M/L$ is constant with redshift (note that a larger
mass threshold is required at higher $z$ for a fixed apparent magnitude
limit).
This type of model generically predicts a steep fall-off in density towards
higher $z$, in contrast with the collisional scenario, and in apparent
disagreement with the newest data  (\cite{steidel:98}) (solid diamonds).
Redshift evolution is
thus a key discriminant between the scenarios. The predictions of the
simple model shown here are similar to those of more detailed
central-quiescent models (\cite{baugh:98,spf:98}), but large uncertainties remain.
Very different results can be obtained if the efficiency of star formation
varies with redshift  (\cite{spf:98}).

\begin{figure}[b!]
\centerline{\psfig{file=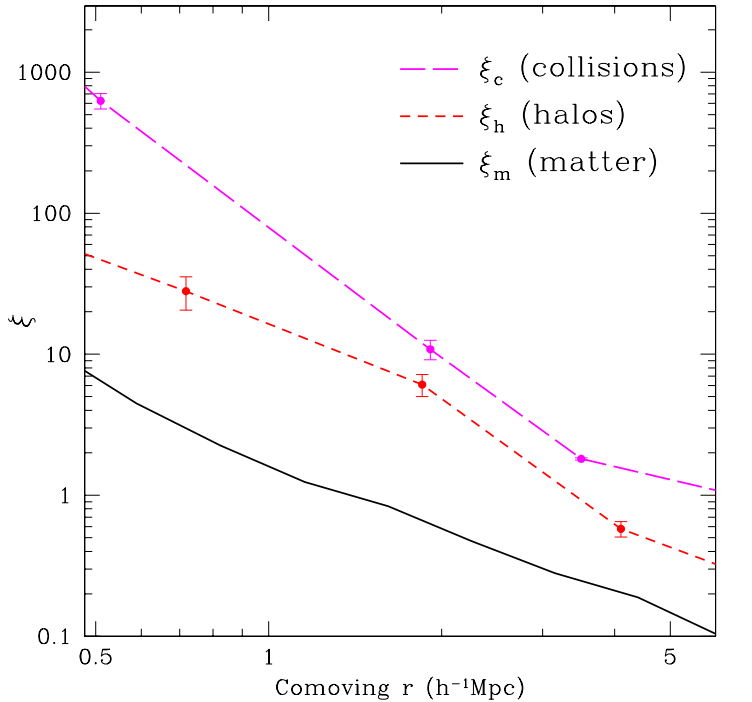
,height=10truecm,width=10truecm}}
\caption{
Two-point autocorrelation functions.
The upper curve is for collisions between DM halos of
$M > 7 \times 10^9 \hsolmass$
that occurred in the redshift interval $3.9 > z > 2.9$.
The error bars are combined Poisson and model-fit ($\Rs$) errors.
The middle curve refers to halos at $z=2.9$ with
$M>10^{12}\hsolmass$, and the lower curve is for the underlying dark matter.
The collisional
starburst scenario thus predicts that LBGs should be strongly
clustered, more than the halos themselves.
}
\label{fig:correlation}
\end{figure}

\section{Clustering properties}

A key observed statistical property of the LBGs is the
strong clustering they manifest.
In the central quiescent model the number density of LBGs sets
their mass scale.  Analytic approximations (Mo et al. 1999)
can then be used in order to derive their clustering properties.
Here, since the collisions are selected by their dynamics, one
must calculate clustering properties directly from the
simulations.

Figure \ref{fig:correlation} depicts the correlation function of the
collisions.
In the range $1-5 \hmpc$, it can be approximated by a power law,
$\xi_{\rm c}(r) \simeq (r/r_0)^{- \gamma}$, with $r_0\simeq 5
\hmpc$ and $\gamma \simeq 2.6$.
Shown for comparison is $\xi_{\rm h}(r) \simeq (r/3.5 h^{-1}
{\rm Mpc})^{-2.2}$ for halos $>10^{12} \hsolmass$; these are what
the central quiescent scenario would identify with LBGs.
Given current uncertainties,
both correlation functions are consistent with the parameters derived from
observations (\cite{adelberger:98,giavalisco:98})
for the simulated cosmology:  $r_0 \simeq 6\hmpc$ and $\gamma \simeq 2$.
Also shown for reference is $\xi_{\rm m}(r)$ of the underlying dark matter.
The relative biases at $5 \hmpc$ are
$(\xi_{\rm c}/\xi_{\rm m})^{1/2} \simeq 3$ and
$(\xi_{\rm h}/\xi_{\rm m})^{1/2} \simeq 2$.
We find that $\sim 30\%$ of the collisions at $z\sim 3$ occur within halos
of mass $\simgreat 10^{12.5}\hsolmass$,
and that most of the rest occur in dense environments near such halos.
This
is consistent with the fact that
the collisions are highly correlated and is confirmed by the
similarity between the auto-correlation of collisions and the
cross-correlation of collisions and halos of mass $>10^{13.0}\hsolmass$
(at $0.5<r<5 \hmpc$).
The combination of gravitational collapse
and a collision rate proportional to the square of the halo number density
can explain why collisions occur mainly near the most massive halos.
A visual confirmation of this trend can be gathered from Figure 1.

\section{Conclusions}

The results presented here provide the first quantitative
results on the clustering of colliding halos at high redshift
and a much more accurate measure of their number density than
earlier semi-analytic
calculations (Somerville et al. 1999), which these new results
generally confirm.  In particular, they show
that starbursts associated with collisions of
relatively low mass halos are consistent with the observed number
density
and clustering of bright LBGs at $z\sim2.5-4.5$, at least in the
$\Lambda$CDM cosmology; other popular cosmologies
should be qualitatively similar~ (\cite{spf:98,wechsler:98}).
Finally, they predict that a key test between models is the
number of LBGs vs. redshift, which falls off much
faster in the central quiescent scenario
than for collisional starbursts.  In addition to further tests,
such as the luminosity function and virial mass measurements, this
should ultimately distinguish between these scenarios.

\acknowledgements
The simulations were performed at NRL
and NCSA.  This work was supported by grants from NASA and NSF at
UCSC
and NMSU, and by Israel Science Foundation and US-Israel
Binational
Science Foundation grants.  P.J. acknowledges a Sweden-America
Foundation Graduate Fellowship, and J.R.P. acknowledges a
Forchheimer
Visiting Professorship at Hebrew University.
We thank Stephan Charlot for letting us use GISSEL98.

\bibliography{mnrasmnemonic,all_refs}

\end{document}